\documentclass[]{aastex7}

\usepackage{amsmath,bm,newtxtext,newtxmath}
\usepackage[T1]{fontenc}
\usepackage[encapsulated]{CJK}

\begin{document}
\begin{CJK*}{UTF8}{gbsn}
\title{Spherically Symmetric Accretion with Self-Gravity: Analytical Formulae and Numerical Validation}

\author[0000-0002-7663-7900,gname='承亮',sname='焦']{Cheng-Liang Jiao (焦承亮)}
\email{jiaocl@ynao.ac.cn}
\affiliation{Yunnan Observatories, Chinese Academy of Sciences, 396 Yangfangwang, Guandu District, Kunming, 650216, P. R. China}
\affiliation{Center for Astronomical Mega-Science, Chinese Academy of Sciences, 20A Datun Road, Chaoyang District, Beijing, 100012, P. R. China}
\affiliation{Key Laboratory for the Structure and Evolution of Celestial Objects, Chinese Academy of Sciences, 396 Yangfangwang, Guandu District, Kunming, 650216, P. R. China}
\correspondingauthor{Cheng-Liang Jiao (焦承亮)}
\email{jiaocl@ynao.ac.cn}
\author{Er-gang Zhao}
\email{zergang@ynao.ac.cn}
\affiliation{Yunnan Observatories, Chinese Academy of Sciences, 396 Yangfangwang, Guandu District, Kunming, 650216, P. R. China}
\affiliation{Key Laboratory for the Structure and Evolution of Celestial Objects, Chinese Academy of Sciences, 396 Yangfangwang, Guandu District, Kunming, 650216, P. R. China}
\author[0000-0002-0796-7009]{Liying Zhu}
\email{zhuly@ynao.ac.cn}
\affiliation{Yunnan Observatories, Chinese Academy of Sciences, 396 Yangfangwang, Guandu District, Kunming, 650216, P. R. China}
\affiliation{University of Chinese Academy of Sciences, No. 1 Yanqihu East Rd, Huairou District, 101408, Beijing, Peopleʼs Republic of China}
\affiliation{Key Laboratory for the Structure and Evolution of Celestial Objects, Chinese Academy of Sciences, 396 Yangfangwang, Guandu District, Kunming, 650216, P. R. China}
\author[0000-0002-5038-5952]{Xiang-dong Shi}
\email{sxd@ynao.ac.cn}
\affiliation{Yunnan Observatories, Chinese Academy of Sciences, 396 Yangfangwang, Guandu District, Kunming, 650216, P. R. China}
\affiliation{Key Laboratory for the Structure and Evolution of Celestial Objects, Chinese Academy of Sciences, 396 Yangfangwang, Guandu District, Kunming, 650216, P. R. China}

\begin{abstract}
Spherically symmetric accretion incorporating self-gravity constitutes a three-point boundary value problem (TPBVP) governed by constraints at the outer boundary, sonic point, and accretor surface. Previous studies have two limitations: either employing an incorrect formula for self-gravity potential in analytical treatments, or introducing additional input parameters in numerical implementations to circumvent solving the full TPBVP. To address these issues, we present a self-consistent TPBVP formulation, solved using the relaxation method. We also derive approximate analytical formulae that enable rapid estimates of self-gravity effects. Our analysis identifies a dimensionless parameter $\beta \equiv 2G \bar{\rho} r_\mathrm{out}^2/a_\mathrm{out}^2$ that characterizes the strength of self-gravity, where $\bar{\rho}$ and $r_\mathrm{out}$ are the mean density and outer radius of the flow, respectively, and $a_\mathrm{out}$ is the adiabatic sound speed of the external medium. For practical estimation, $\bar{\rho}$ may be approximated by the external medium density $\rho_\mathrm{out}$. We identify an upper limit for $\beta$, beyond which steady accretion becomes unsustainable---a behavior consistent with classical gravitational instability that previous studies failed to capture.
The accretion rate enhancement decreases monotonically as the adiabatic index $\gamma$ increases. For $\gamma=5/3$, self-gravity ceases to augment the accretion rate.
These theoretical predictions are validated by our numerical solutions.
We further apply our results to two astrophysical scenarios: hyper-Eddington accretion onto supermassive black hole seeds in the early Universe, where self-gravity is significant; and accretion onto stellar-mass objects embedded in active galactic nuclei (AGN) disks, where self-gravity is non-negligible under certain conditions and should be evaluated using $\beta$.
\end{abstract}
%
\section{Introduction}\label{intro}
Steady, spherically symmetric accretion onto a central accretor was first systematically explored by \cite{Bondi1952}, who demonstrated the existence of a maximum accretion rate under specified conditions of the mass of the accretor and the density and sound speed of the ambient medium. The critical rate corresponds to the unique transonic accretion solution (type \uppercase\expandafter{\romannumeral2} solution in \citealt{Bondi1952}), and is physically preferred because it minimizes the energy of the system and ensures that all flow variables exhibit smooth and uniform behavior.
This theoretically derived accretion rate, commonly known as the Bondi accretion rate (denoted $\dot{M}_{\mathrm{B}}$), has become a fundamental tool for estimating accretion rates in diverse astrophysical systems ranging from protostars to supermassive black holes (SMBHs).
However, the classical Bondi model is a relatively simple model and some of its assumptions may become inadequate in certain astrophysical scenarios. A notable limitation arises from its neglect of self-gravity of the accretion flow, which becomes important in gas-rich environments.
For example, \cite{Kohei2016} found that steady hyper-Eddington accretion ($\dot{M} \simeq \dot{M}_{\mathrm{B}} \gtrsim 5000 L_{\mathrm{Edd}}/c^2$, where $L_{\mathrm{Edd}}$ is the Eddington luminosity) can occur for spherically symmetric accretion flows onto black holes (BHs) embedded in dense metal-poor clouds, which provides an explanation for the rapid formation of SMBHs ($>10^9M_\odot$) in the early Universe (redshift $z>6$; e.g., \citealt{Wu2015,Yang2020,WFG2021}).
Such extreme accretion rates are enabled by the high density of the ambient gas, a condition under which the self-gravity of the flow can no longer be ignored. 
Another particularly relevant case is the accretion by stellar-mass  objects (SMOs) embedded in the disks of active galactic nuclei (AGNs), where the density of the external medium reaches $\sim 10^{14}$ cm$^{-3}$ \citep{Kato2008, Wang2021}.
For a $10M_\odot$ object corotating with the AGN disk, the gas mass enclosed within its Bondi radius can reach $\sim 240M_\odot$ [see equation (6) of \citealt{Wang2021}], significantly exceeding the mass of the accretor. While the Bondi accretion rate or the interpolation formula between that and the Hoyle-Lyttleton accretion \citep{HL1939,Shima1985,Edgar2004} is commonly adopted in related studies \citep[e.g.,][]{Cant2021,Wang2021,Wang2021b,Chen2023}, the omission of self-gravity could introduce systematic biases in accretion rate estimates. These considerations necessitate the formulation of an improved model that accounts for the self-gravity of the accretion flow.

The inclusion of self-gravity in steady spherical accretion was first explored by \cite{Chia1978}, who reported significant accretion rate enhancement due to self-gravity. Their analysis, however, employed an incorrect formula for the gravitational potential of the accretion flow (see Section \ref{ana_eqs}), casting doubt on the reliability of their results.
Subsequent numerical investigations by \cite{Datta2016} and \cite{Moham2018} introduced ad hoc parameters ($s$ and $1/m_*$, respectively) to characterize the mass ratio between the accretion flow and the central object. This approach fundamentally alters the mathematical structure of the problem by artificially prescribing mass boundary conditions (BCs) at the outer boundary, thereby avoiding the inherent three-point boundary value problem (TPBVP) that naturally arises in self-gravitating systems. From a mathematical point of view, the self-gravity at radius $r$ depends on the enclosed mass of the accretion flow inside $r$ (denoted $M_r$), which introduces a new dependent variable $M_r$. This necessitates an additional BC, which is naturally provided at the surface of the central accretor where $M_r$ becomes 0. The transonic accretion solution is thus uniquely determined by BCs at three different locations: the outer radius, the sonic radius, and the surface of the accretor, constituting a TPBVP.
While the parametric approximations in \cite{Datta2016} and \cite{Moham2018} simplify computations by reducing it to a two-point problem, they impose unphysical constraints on the total mass of the flow, which should emerge self-consistently from BCs.
Directly solving the TPBVP is therefore essential to eliminate artificial mass scaling assumptions that limit predictive power across astrophysical regimes. Concurrently, given the mathematical challenges inherent in solving TPBVPs, deriving approximate analytical formulae that capture self-gravity effects is equally imperative. Such approximations, once verified against the numerical solutions of the TPBVP, would enable rapid estimates of self-gravity effects.

In this paper, we develop a self-consistent model for steady, spherically symmetric accretion with self-gravity. Unlike previous approaches, we solve the full TPBVP numerically without introducing additional artificial parameters. We also derive analytical formulae for the accretion rate and sonic radius, based on a dimensionless parameter that characterizes the strength of self-gravity. We then compare numerical solutions of the TPBVP with our analytical formulae and those of \citet{Chia1978}, validating our formulae. Finally, we apply our model to two realistic astrophysical scenarios: accretion onto SMBH seeds in the early Universe, and onto SMOs embedded in AGN disks. These scenarios highlight the astrophysical relevance of our findings.
The remainder of this paper is organized as follows. We present the TPBVP formulation of the problem in Section \ref{model}. We then derive approximate analytical formulae in Section \ref{ana_eqs} and an upper limit for self-gravity effects in Section \ref{upperlimit}. Numerical solutions are presented and compared with analytical predictions in Section~\ref{results}. 
Astrophysical applications of our model are explored in Section~\ref{apps}.
Our findings are summarized and discussed in Section \ref{summary}.
\section{Model Formulation}\label{model}
\subsection{Basic Equations and Assumptions}\label{eqs}
We consider a steady, spherically symmetric accretion flow onto a central accretor of mass $M_*$. The continuity and radial momentum equations are given by
\begin{equation}\label{eq_cont}
	\frac{\mathrm{d}}{\mathrm{d} r}(r^2\rho v_r)=0,
\end{equation}
\begin{equation}\label{eq_r_motion2}
	v_r\frac{\mathrm{d} v_r}{\mathrm{d} r} + \frac{1}{\rho} \frac{\mathrm{d} p}{\mathrm{d} r} + \frac{GM_\mathrm{t}}{r^2}=0,
\end{equation}
where $\rho$, $v_r$, and $p$ denote density, radial velocity, and pressure, respectively, and $M_\mathrm{t}$ represents the total mass inside radius $r$. Equation (\ref{eq_r_motion2}) is applicable because the gravitational force exerted by the mass outside radius $r$ cancels out due to Newton's shell theorem. The mass of the accretion flow inside radius $r$ is
\begin{equation}\label{eq_Mr}
	M_r=\int_{r_*}^r 4 \pi r'^2 \rho \mathrm{d} r',
\end{equation}
where $r_*$ is the accretor surface. Thus we get $M_\mathrm{t}=M_r+M_*$. 
We expect that the mass accreted during the time interval of interest is small compared to $M_*$ and neglect the change of $M_*$. Differentiating $M_\mathrm{t}$ with respect to $r$ yields
\begin{equation}\label{eq_Mt}
	\frac{\mathrm{d} M_\mathrm{t}}{\mathrm{d} r}=4 \pi \rho r^2
\end{equation}
for $r \geq r_*$, which can also be derived from Gauss's law for gravity.
Following the classical Bondi model, we assume that the flow is adiabatic, such that the energy equation can be written as 
\begin{equation}\label{eq_E}
	p\rho^{-\gamma}=\mathrm{const},
\end{equation}
where $\gamma$ is the adiabatic index and assumed to be constant. We consider $1 \leq \gamma \leq 5/3$.
%
\subsection{Boundary Conditions}
Equations (\ref{eq_cont}), (\ref{eq_r_motion2}), (\ref{eq_Mt}), and (\ref{eq_E}) constitute a system comprising three ordinary differential equations (ODEs) and an algebraic equation with an unknown constant parameter (which can be rewritten as an ODE), governing four dependent variables. Solving this system requires four BCs. 
At the outer boundary ($r=r_\mathrm{out}$), we impose asymptotic convergence to the ambient medium:
\begin{equation}\label{OBC1}
	r=r_\mathrm{out}:\quad \rho=\rho_\mathrm{out},
\end{equation}
\begin{equation}\label{OBC2}
	r=r_\mathrm{out}:\quad a=a_\mathrm{out},
\end{equation}
where $a$ is the adiabatic sound speed, defined as
\begin{equation}\label{eq_a}
	a \equiv \sqrt{\frac{\mathrm{d} p}{\mathrm{d} \rho}}=\sqrt{\frac{\gamma p}{\rho}}.
\end{equation}
The adiabatic sound speed is determined by temperature $T$ according to the ideal gas law\footnote{Strictly speaking, Equation~\eqref{eq_T} is valid only when gas pressure dominates the total pressure. Other pressure components, such as radiation or magnetic pressure, may become important under certain conditions, in which case $p$ should be interpreted as the total pressure. This, however, does not affect our derivation, which does not rely on Equation~\eqref{eq_T}.}
\begin{equation}\label{eq_T}
	a=\sqrt{\frac{\gamma R_\mathrm{gas} T}{\mu}},
\end{equation}
where $R_\mathrm{gas}$ is the gas constant and $\mu$ the mean molecular weight of the gas, so the outer BCs essentially describe the density and temperature of the ambient medium. One may either use $a$ as a dependent variable in place of $p$, or calculate $p_\mathrm{out}$ with Equation (\ref{eq_a}).

By combining with other equations, Equation (\ref{eq_r_motion2}) can be rewritten as
\begin{equation}\label{eq_dvr}
	\frac{1}{2}\left(1-\frac{a^2}{v_r^2}\right) \frac{\mathrm{d}v_r^2}{\mathrm{d} r}=\frac{2a^2}{r}-\frac{G M_\mathrm{t}}{r^2},
\end{equation}
which reduces to that used in the classical Bondi model [e.g., equation (2.27) in \citealt{Frank2002}] in the absence of self-gravity ($M_\mathrm{t}=M_*$).
As analyzed by \cite{Frank2002} (see also \citealt{Holzer1970}), six types of solutions exist depending on BC configurations. Our focus is the transonic accretion solution, which is subsonic at $r_\mathrm{out}$, supersonic near $r_*$ (when $r_*$ is not larger than the expected sonic radius), and crosses the sonic point smoothly. This requires that the numerator and denominator of ${\mathrm{d}v_r^2}/{\mathrm{d} r}$ in Equation (\ref{eq_dvr}) vanish simultaneously at the sonic radius $r_\mathrm{c}$, yielding
\begin{equation}\label{D1}
	r=r_\mathrm{c}:\quad {v_r}^2=a^2,
\end{equation}
\begin{equation}\label{N1}
	r=r_\mathrm{c}:\quad \frac{2a^2}{r}=\frac{G M_\mathrm{t}}{r^2},
\end{equation}
which provide two BCs and introduce $r_\mathrm{c}$ as an additional parameter to solve for, equivalent to one BC.
The final BC is naturally imposed at the accretor surface,
\begin{equation}\label{eq_IBC}
	r=r_*:\quad M_\mathrm{t}=M_*.
\end{equation}
The problem is thus a TPBVP, which can be solved with the relaxation method \citep{Press2002}.

\section{Analytical formulae}\label{asol}
We derive approximate analytical formulae for the model in this section.
While these formulae are less precise than the numerical solutions presented in Section \ref{results}, they provide key physical insights into the problem and enable rapid estimates of self-gravity effects, bypassing the computational complexity of TPBVPs. Specifically, we define the "self-gravity parameter" and derive expressions for the sonic radius and accretion rate based on this parameter in Section \ref{ana_eqs}. We then establish an upper limit for self-gravity effects in Section \ref{upperlimit}, beyond which a steady, spherically symmetric model no longer applies. These approximate formulae will be compared with numerical solutions in Section \ref{results}.
\subsection{Self-Gravity Parameter and Formulae for Accretion Rate and Sonic Radius}\label{ana_eqs}
The problem can be analyzed using the integral of Equation (\ref{eq_r_motion2}), commonly known as the Bernoulli integral, as in the classical Bondi model. Here we consider a realistic astrophysical system that is suitable for the steady, spherically symmetric accretion model. 
We expect the length scale of the external medium to be much larger than the Bondi radius $r_\mathrm{B}$ \citep[also known as the accretion radius;][]{Frank2002},
\begin{equation}\label{rB}
	r_\mathrm{B} \equiv \frac{2G M_\mathrm{*}}{a_\mathrm{out}^2}.
\end{equation}
If the size of the external medium is comparable to $r_\mathrm{B}$, then the accretion would cause the medium to shrink, making it difficult to maintain a steady flow. Thus we set $r_\mathrm{out} \gg r_\mathrm{B}$. 
On the other hand, $r_\mathrm{out}$ cannot be arbitrarily large, or the medium would collapse due to gravitational instability \citep{Ebert1955,Bonnor1956,book2017}. The critical radius beyond which the collapse would happen can be estimated with \citep{Bonnor1956}
\begin{equation}\label{rgi}
	r_\mathrm{gi} \approx 0.76\sqrt{\frac{R T}{\mu G \rho_\mathrm{out}}}
    =\frac{0.76a_\mathrm{out}}{\sqrt{\gamma G \rho_\mathrm{out}}},
\end{equation}
where $\rho_\mathrm{out}$ is taken as the mean density of the medium.\footnote{Strictly speaking, this formula is derived for isothermal gas. It does agree with the limitation derived from our model, as shown in Section \ref{upperlimit}.}
Therefore, we generally have $r_\mathrm{B} \ll r_\mathrm{out} \leq r_\mathrm{gi}$ in a realistic case.
The mass of the gas enclosed within $r_\mathrm{B}$ can be approximated as $M_r(r_\mathrm{B}) \sim {4\pi}\rho_\mathrm{out} r_\mathrm{B}^3/3$. Its ratio to the mass of the central accretor $M_*$ can thus be estimated as
\begin{equation}\label{ratio2}
    \frac{M_r(r_\mathrm{B})}{M_*} \sim \frac{4\pi\rho_\mathrm{out} r_\mathrm{B}^3 }{3M_*}=\frac{8\alpha}{3},
\end{equation}
where $\alpha$ is the "self-gravitational parameter"  proposed by \citet{Chia1978}, expressed in our notation as
\begin{equation}\label{alpha}
    \alpha=\frac{\pi\rho_\mathrm{out}r_\mathrm{B}^3}{2M_*}.
\end{equation}
Meanwhile, $\alpha$ itself is proportional to $(r_\mathrm{B}/r_\mathrm{gi})^2$:
\begin{equation}\label{ratio1}
    \left(\frac{r_\mathrm{B}}{r_\mathrm{gi}}\right)^2 \approx \frac{4\gamma G^3M_*^2\rho_\mathrm{out}}{0.76^2a_\mathrm{out}^6}\approx\frac{1.73\gamma}{\pi}\cdot \alpha.
\end{equation}
Since the coefficients of $\alpha$ in both Equations~\eqref{ratio2} and \eqref{ratio1} are on the order of unity, we find that to leading order, $M_r(r_\mathrm{B})/M_* \sim \alpha \sim (r_\mathrm{B}/r_\mathrm{gi})^2$.
The main result of \cite{Chia1978} is that the enhancement of the accretion rate due to self-gravity is dependent on $\alpha$. This will be checked in subsequent derivations.

While we do not know the mass distribution of the accretion flow, it can be estimated based on characteristic length scales. Combining Equation (\ref{ratio1}) with $r_\mathrm{B} \ll  r_\mathrm{gi}$,  we deduce that realistic systems satisfy $\alpha \ll 1$.
We expect that the magnitude of radial velocity increases as $r$ decreases, so that the density profile should not be steeper than $\rho \propto r^{-2}$, which corresponds to a flat profile of $v_r$ according to the continuity equation. The mass of the accretion flow is thus mostly distributed in the outer regions, since the integrand for mass, $4\pi r^2\rho$, increases with $r$. With $r_\mathrm{B} \ll r_\mathrm{out}$, we deduce that $M_r(r_\mathrm{B}) \ll M_r(r_\mathrm{out})$.

For $\gamma \neq 1$ (the $\gamma=1$ case will be discussed later), the Bernoulli integral yields
\begin{equation}\label{eq_Be}
	\frac{v_\mathrm{c}^2}{2}+\frac{a_\mathrm{c}^2}{\gamma-1}+\Psi_*(r_\mathrm{c})+\Psi_\mathrm{sg}(r_\mathrm{c})=\frac{v_\mathrm{out}^2}{2}+\frac{a_\mathrm{out}^2}{\gamma-1}+\Psi_*(r_\mathrm{out})+\Psi_\mathrm{sg}(r_\mathrm{out}),
\end{equation}
where $v_\mathrm{out}$ is the radial velocity at the outer boundary ($v_\mathrm{out}^2/2$ is typically negligible compared to other terms), $v_\mathrm{c}$ and $a_\mathrm{c}$ are the radial velocity and sound speed at the sonic radius, and $\Psi_*$ ($=-GM_*/r$) and $\Psi_\mathrm{sg}$ are the gravitational potential of the central accretor and the accretion flow, respectively. By Newton's shell theorem, the gravitational potential at radius $r$ imposed by a shell at radius $R$ with mass $\Delta M=4\pi\rho R^2\Delta R$ is
\begin{equation}
\Delta\Psi = 
\begin{cases}
-\frac{G\Delta M}{r}, & r>R, \\
-\frac{G\Delta M}{R}, & r\leq R. \\
\end{cases}
\end{equation}
The total self-gravity potential is thus $\Psi_\mathrm{sg}=\Psi_1+\Psi_2$, where
\begin{equation}
\Psi_1=-\frac{4\pi G}{r}\int_{r_*}^{r}\rho R^2 \mathrm{d}R=-\frac{GM_r}{r}
\end{equation}
describes the potential from mass interior to $r$, and
\begin{equation}\label{eq_phi2}
\Psi_2=-{4\pi G}\int_{r}^{r_\mathrm{out}}\rho R \mathrm{d}R
\end{equation}
represents the contribution from mass between $r$ and $r_\mathrm{out}$. We note that the potential imposed by mass outside $r_\mathrm{out}$ appears on both sides of Equation (\ref{eq_Be}) and always cancels out, so that it can be neglected. 
\cite{Chia1978} further neglected $\Psi_2$, $\Psi_*(r_\mathrm{out})$, and $\Psi_\mathrm{sg}(r_\mathrm{out})$ [see their equations (2.4) and (2.8)].
While $\Psi_*(r_\mathrm{out})$ is generally negligible due to $r_\mathrm{out} \gg r_\mathrm{B} >r_\mathrm{c}$, the other two terms are not:
$\Psi_2$ dominates the self-gravity potential at small radii (e.g., the sonic radius; see derivation below), and $\Psi_\mathrm{sg}(r_\mathrm{out})$ represents the total self-gravity potential at the outer boundary. 
Since all $r_\mathrm{out}$-dependent terms in the Bernoulli integral are omitted, their predictions of self-gravity effects become independent of $r_\mathrm{out}$, contrary to expectations from classical gravitational instability (see Section~\ref{mdot_vs_beta} for details).

At the outer boundary and the sonic radius, we get
\begin{equation}\label{sgrout0}
\Psi_\mathrm{sg}(r_\mathrm{out})=-\frac{GM_r(r_\mathrm{out})}{r_\mathrm{out}}
\end{equation}
\begin{equation}\label{sgrc0}
\Psi_\mathrm{sg}(r_\mathrm{c})=-\frac{GM_r(r_\mathrm{c})}{r_\mathrm{c}}-{4\pi G}\int_{r_\mathrm{c}}^{r_\mathrm{out}}\rho R \mathrm{d}R,
\end{equation}
respectively.
The Bondi solution has shown that outside $r_\mathrm{B}$, gas density becomes almost constant since the effects of gravity there are weak. Here we consider cases not significantly deviating from the Bondi solution, so that the density outside $r_\mathrm{B}$ may still be approximated as constant (see Section \ref{results} for further discussion). Combined with the fact that the mass of the flow is mostly distributed in the outer regions, we can approximately treat the density as a constant value of $\bar{\rho}$ in the mass integrals. Defining the total mass of the accretion flow as $M_\mathrm{flow} \equiv M_r(r_\mathrm{out})$, the mean density $\bar{\rho}$ can be expressed as
\begin{equation}\label{rho_mean}
\bar{\rho} = \frac{3M_\mathrm{flow}}{4\pi r_\mathrm{out}^3}.
\end{equation}
With Equations (\ref{sgrout0}), (\ref{sgrc0}), and (\ref{rho_mean}), we obtain
\begin{equation}\label{sgrout}
\Psi_\mathrm{sg}(r_\mathrm{out}) = -\frac{GM_\mathrm{flow}}{r_\mathrm{out}},
\end{equation}
\begin{equation}\label{sgrc}
\Psi_\mathrm{sg}(r_\mathrm{c}) \approx -\frac{3GM_\mathrm{flow}}{2r_\mathrm{out}},
\end{equation}
where $\Psi_1(r_\mathrm{c})=-{GM_r(r_\mathrm{c})}/{r_\mathrm{c}}$ is neglected because the variation in $\rho$ cannot be steeper than $r^{-2}$ and $r_\mathrm{out} \gg r_\mathrm{c}$.
Furthermore, $\alpha \ll 1$ implies that the mass of the accretion flow inside $r_\mathrm{B}$ is much smaller than $M_*$. Since $r_\mathrm{B} > r_\mathrm{c}$, the mass of the flow inside $r_\mathrm{c}$ is even smaller, so that we can set $M_t(r_\mathrm{c}) \approx M_*$ in Equation (\ref{N1}), which will be verified in Section \ref{results}. Equations (\ref{D1}) and (\ref{N1}) then yield
\begin{equation}\label{eq_DN1}
	{v_\mathrm{c}}^2=a_\mathrm{c}^2=\frac{G M_*}{2r_\mathrm{c}}.
\end{equation}
Substituting Equations (\ref{rB}), (\ref{sgrout}), (\ref{sgrc}), and (\ref{eq_DN1}) into Equation (\ref{eq_Be}), and defining $x_\mathrm{c} \equiv r_\mathrm{c}/r_\mathrm{B}$, we derive
\begin{equation}\label{eq_xc}
	x_\mathrm{c} = \frac{3(5-3\gamma)}{12\left[2-(\gamma-1)\frac{r_\mathrm{B}}{r_\mathrm{out}}\right]+
    \frac{8\pi(\gamma-1)\bar{\rho} r_\mathrm{out}^2 r_\mathrm{B}}{M_*}}.
\end{equation}
Introducing the dimensionless ratios $x_\mathrm{out} \equiv r_\mathrm{out}/r_\mathrm{B}$ and 
\begin{equation}\label{eq_beta}
	\beta \equiv \frac{\bar{\rho} r_\mathrm{out}^2 r_\mathrm{B}}{M_*} = \frac{2G\bar{\rho} r_\mathrm{out}^2}{a_\mathrm{out}^2},
\end{equation}
Equation (\ref{eq_xc}) simplifies to
\begin{equation}\label{eq_xc3}
	x_\mathrm{c} = \frac{3(5-3\gamma)}{12\left[2-(\gamma-1)/{x_\mathrm{out}}\right]+
    8\pi(\gamma-1)\beta}.
\end{equation}
As we expect $x_\mathrm{out} \gg 1$, the location of the sonic point $x_\mathrm{c}$ is primarily governed by $\gamma$ and $\beta$, while $\beta=0$ describes the case that neglects self-gravity. Thus, we 
define $\beta$ as the self-gravity parameter in this paper.
The $x_\mathrm{out}$ term quantifies the shift of the sonic radius when the outer boundary is set at a finite radius, instead of at infinity as in the classical Bondi model \citep{Samadi2019}. We can estimate the value of $\beta$ by setting $\bar{\rho} \approx \rho_\mathrm{out}$, such that $\beta \approx \tilde{\beta} = \rho_\mathrm{out} r_\mathrm{out}^2 r_\mathrm{B}/{M_*} = 2\alpha x_\mathrm{out}^2/\pi$. It can also be expressed solely in terms of the properties of the external medium: $\tilde{\beta} = 2G \rho_\mathrm{out} r_\mathrm{out}^2 /a_\mathrm{out}^2$. In contrast to the self-gravity parameter $\alpha$ in \cite{Chia1978}, which scales with $M_*^2$ due to $r_\mathrm{B} \propto M_*$, our description is independent of $M_*$. Consequently, the self-gravity effects depend exclusively on the properties of the external medium, specifically the density $\rho_\mathrm{out}$, the adiabatic sound speed $a_\mathrm{out}$, the outer radius $r_\mathrm{out}$, and the adiabatic index $\gamma$.
We note that while $\beta$ is independent of $M_*$, the condition $x_\mathrm{out} \gg 1$ implicitly constrains the relation between $M_\mathrm{flow}$ and $M_*$. We can rewrite $\beta$ as
\begin{equation}
	\beta = \frac{3}{4\pi x_\mathrm{out}}\frac{M_\mathrm{flow}}{M_*}.
\end{equation}
Since $x_\mathrm{out} \gg 1$, we derive $M_\mathrm{flow} \gg \beta M_* $.

The accretion rate can be calculated as $\dot{M}=4\pi r_\mathrm{c}^2\rho_\mathrm{c}|v_\mathrm{c}|$. From Equation (\ref{eq_DN1}), we know that $|v_\mathrm{c}|=a_\mathrm{c} \propto r_\mathrm{c}^{-1/2}$. From Equations (\ref{eq_E}) and (\ref{eq_a}), we obtain $\rho_\mathrm{c}\propto a_\mathrm{c}^{2/(\gamma-1)}\propto r_\mathrm{c}^{-1/(\gamma-1)}$. Thus we know that $\dot{M} \propto r_\mathrm{c}^{3/2-1/(\gamma-1)}=r_\mathrm{c}^{(3\gamma-5)/[2(\gamma-1)]}$. The exact formula can be written as
\begin{equation}\label{eq_mdot}
    \dot{M}=\pi \lambda_\mathrm{sg} G^2 M_*^2 \rho_\mathrm{out}a_\mathrm{out}^{-3},
\end{equation}
where $\lambda_\mathrm{sg}$ is a dimensionless parameter defined as
\begin{equation}\label{eq_lam}
    \lambda_\mathrm{sg} \equiv (4x_\mathrm{c})^{(3\gamma-5)/[2(\gamma-1)]}.
\end{equation}
For consistency checks, we compare with the classical Bondi model, which neglects self-gravity ($\beta=0$) and sets the outer radius at infinity ($x_\mathrm{out} \to \infty$). In this case, $x_\mathrm{c}=(5-3\gamma)/8$, and the accretion rate becomes
\begin{equation}\label{eq_mdot_b}
    \dot{M}_\mathrm{B}=\pi  G^2 M_*^2 \rho_\mathrm{out}a_\mathrm{out}^{-3} \left(\frac{5-3\gamma}{2}\right)^{\frac{3\gamma-5}{2 (\gamma -1)}},
\end{equation}
in agreement with the standard Bondi solution \citep{Frank2002}. To quantify self-gravity effects, we define the normalized accretion rate
\begin{equation}\label{eq_mdot2}
    \dot{m}\equiv \frac{\dot{M}}{\dot{M}_\mathrm{B}}=\left[\frac{1}{3} \pi  \beta  (\gamma -1)-\frac{\gamma -1}{2 x_\mathrm{out}}+1\right]^{\frac{5-3 \gamma }{2 (\gamma -1)}},
\end{equation}
which measures the enhancement of the accretion rate due to self-gravity.
For $1<\gamma<5/3$, $x_\mathrm{c}$ decreases and  $\dot{m}$ increases as $\beta$ increases, indicating that self-gravity shifts the sonic point inward and enhances accretion.
For $\gamma=5/3$, $x_\mathrm{c}$ remains zero and $\dot{m}$ is unaffected by self-gravity under our approximations.
For $\gamma=1$, we have $p\propto \rho$ and $a=\sqrt{p/\rho}=\mathrm{const}$. Using Equations (\ref{D1}) and (\ref{N1}) under the approximation $M_t(r_\mathrm{c}) \approx M_*$, we obtain $x_\mathrm{c}=1/4$.
The Bernoulli integral becomes
\begin{equation}\label{eq_Be2}
	\frac{v_\mathrm{c}^2}{2}+a_\mathrm{out}^2\ln{\rho_\mathrm{c}}+\Psi_*(r_\mathrm{c})+\Psi_\mathrm{sg}(r_\mathrm{c})=\frac{v_\mathrm{out}^2}{2}+a_\mathrm{out}^2\ln{\rho_\mathrm{out}}+\Psi_*(r_\mathrm{out})+\Psi_\mathrm{sg}(r_\mathrm{out}),
\end{equation}
where $\rho_\mathrm{c}$ is the density at $r_\mathrm{c}$. Following a similar derivation, we find
\begin{equation}\label{eq_rhoc}
	\ln{\rho_\mathrm{c}}=\ln{\rho_\mathrm{out}}+\frac{3}{2}-\frac{1}{2x_\mathrm{out}}+\frac{\pi\beta}{3}.
\end{equation}
With $x_\mathrm{c}=1/4$, we can derive the formula of $\dot{m}$ for $\gamma=1$, which is
\begin{equation}\label{eq_mdot_gamma1}
    \dot{m}|_{\gamma=1} = \exp{\left(\frac{\pi\beta}{3}-\frac{1}{2x_\mathrm{out}}\right)}.
\end{equation}
The formulae of $x_\mathrm{c}$ and $\dot{m}$ agree with the limits of Equations (\ref{eq_xc3}) and (\ref{eq_mdot2}) as $\gamma \to 1$. Notably, for $\gamma=1$, $x_\mathrm{c}$ remains almost constant while $\dot{m}$ increases with $\beta$. Considering the fact that $M_t(r_\mathrm{c}) > M_*$, Equation (\ref{N1}) indicates a slight increase in $x_\mathrm{c}$ as $\beta$ increases, which is opposite to the trend for larger values of $\gamma$.
\subsection{Upper Limit of Self-Gravity Effects for a Steady Flow}\label{upperlimit}
Equation (\ref{eq_dvr}) can be expressed in dimensionless form as
\begin{equation}\label{eq_dlnvr}
	\left(\frac{v_r^2}{a^2}-1\right) \frac{\mathrm{d}\ln |v_r|}{\mathrm{d}\ln r}=2-\frac{G M_\mathrm{t}}{a^2r}.
\end{equation}
At the outer boundary, the flow typically has $v_r^2 \ll a^2$. 
We also expect that for an accretion flow, $|v_r|$ should increase as $r$ decreases, such that ${\mathrm{d}\ln |v_r|}/{\mathrm{d}\ln r}<0$. Thus we get
\begin{equation}\label{eq_ul1}
	\frac{G M_\mathrm{t}(r_\mathrm{out})}{a^2r_\mathrm{out}}<2.
\end{equation}
With $M_\mathrm{t}(r_\mathrm{out})=M_\mathrm{flow}+M_*$, Equations (\ref{rho_mean}), (\ref{eq_beta}), and (\ref{eq_ul1}) yield
\begin{equation}\label{eq_ulbeta}
	\beta<\frac{3}{\pi}\left(1-\frac{1}{4x_\mathrm{out}}\right),
\end{equation}
which defines an upper limit for $\beta$ (denoted $\beta_\mathrm{max}$).
If $\beta$ exceeds $\beta_\mathrm{max}$, the larger gravity would cause the radial speed to decrease as $r$ decreases, which is physically counter-intuitive. An alternative interpretation is that to ensure the increase of $|v_r|$ with decreasing $r$ beyond $\beta_\mathrm{max}$, the inward flow motion has to be supersonic at the outer boundary, which is also implausible in a steady accretion flow. These contradictions imply that a steady accretion model can no longer describe the flow sufficiently for $\beta > \beta_\mathrm{max}$ (see Section \ref{summary} for further discussion).
Using Equation (\ref{eq_mdot2}), the maximum value of $\dot{m}$ is
\begin{equation}\label{eq_mdotmax}
	\dot{m}_\mathrm{max} = \left[\gamma -\frac{3 (\gamma -1)}{4 x_\mathrm{out}}\right]^{\frac{5-3 \gamma }{2 (\gamma -1)}},
\end{equation}
which becomes $\dot{m}_\mathrm{max} = \exp{[1-3/(4x_\mathrm{out})]}$ for $\gamma=1$. 

\cite{Bonnor1956} found that a critical mass-to-radius ratio exists for a spherical mass of isothermal gas, which is only gravitationally stable when
\begin{equation}\label{eq_ulgi}
	\frac{G M(r)}{a^2r} \leq \frac{2.4}{\gamma}.
\end{equation}
Their result agrees remarkably well with Equation (\ref{eq_ul1}), despite being derived from completely different governing equations. This consistency demonstrates that the existence of an upper limit for self-gravity in steady accretion flows is not a mathematical artifact, but has genuine physical significance.
\section{Numerical Results}\label{results}
We present transonic global solutions obtained by solving the TPBVP formulated in Section \ref{model}, which are compared with the analytical formulae derived in Section \ref{asol}. Here we use the approximate formula for $\beta$,
\begin{equation}\label{eq_beta2}
\tilde{\beta} = \frac{\rho_\mathrm{out} r_\mathrm{out}^2 r_\mathrm{B}}{M_*} = \frac{2G\rho_\mathrm{out} r_\mathrm{out}^2}{a_\mathrm{out}^2},
\end{equation}
because Equation (\ref{eq_beta}) requires the value of $\bar{\rho}$ to evaluate, which only becomes known after a transonic global solution is obtained. Using $\tilde{\beta}$ mimics the situation where we want to evaluate the influence of self-gravity without calculating a full numerical solution. We note that $\beta_\mathrm{max}$ is independent of this approximation and determined by Equation (\ref{eq_ulbeta}), which simplifies to $\beta_\mathrm{max}\approx 3/\pi \approx 0.95$ for $x_\mathrm{out} \gg 1$.

The value of $\gamma$ critically affects solution behavior, especially for $\gamma=5/3$, where both the classical Bondi model and our analytical formula predict a sonic radius at the origin. This effectively prevents a numerical calculation of the BCs at the sonic radius, since both $v_r$ and $a$ approach infinity there. To avoid this singularity, here we show results for $\gamma=1.66$ as a representative case of $\gamma \to 5/3$.

While our formulation requires an inner BC at $r=r_*$, the specific choice of $r_*$ becomes inconsequential when it is sufficiently small ($r_* \ll r_\mathrm{c}$).
This is because the inner BC is effectively satisfied when the mass of the flow between the actual $r_*$ and the adopted $r_*$ in the calculation is negligible compared to $M_*$ (i.e., within the numerical tolerance). 
For simplicity, here we uniformly set $r_*=3GM_*/c^2$, corresponding to the circular photon orbit of a BH. This setting is valid except for extreme cases such as a central supergiant with an unusually large $r_*$.
\subsection{Global Structure of Transonic Solutions}\label{sols}
\subsubsection{Solutions for Different Values of $\tilde{\beta}$}\label{dif_beta}
\begin{figure*}[ht]
    \centering
    \includegraphics[width=\textwidth]{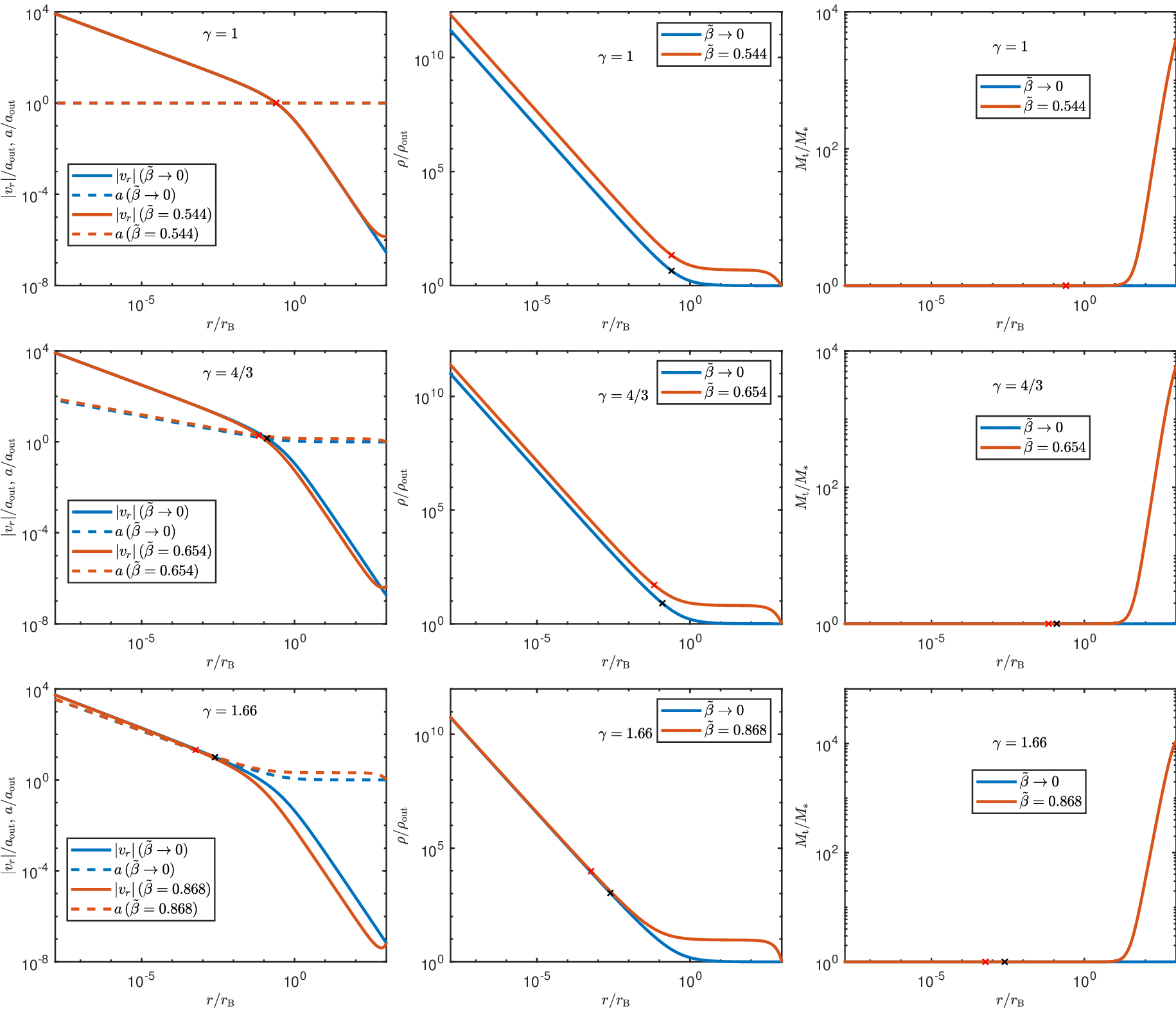}
    \caption{Global solutions of the TPBVP. The left, middle, and right columns show the radial profiles of radial velocity magnitude $|v_r|$ (solid lines) and adiabatic sound speed $a$ (dashed lines), density $\rho$, and total mass $M_\mathrm{t}$, respectively.
The top, middle, and bottom rows correspond to $\gamma=1$, $4/3$, and $1.66$, respectively. The blue and red lines represent solutions with negligible and critical values of $\tilde{\beta}$, where the critical values are determined numerically beyond which the relaxation method no longer converges.
The sonic points are marked by black and red crosses for the negligible and critical $\tilde{\beta}$ cases, respectively. 
Other parameters are 
$a_\mathrm{out}=10^{-4}c$, $x_\mathrm{out}=1000$, and $M_*=M_\odot$.}
    \label{fig1}
\end{figure*}

We show the global structure of transonic solutions in Figure \ref{fig1}, calculated with $a_\mathrm{out}=10^{-4}c$ (corresponding to $T_\mathrm{out}$ on the order of $10^{4}$K), $x_\mathrm{out}=1000$, and $M_*=M_\odot$. Three values of $\gamma$ are adopted, specifically $\gamma=1$, $4/3$, and $1.66$, corresponding to the top, middle, and bottom rows in Figure \ref{fig1}. Two solutions are presented for each $\gamma$: one is calculated with $\rho_\mathrm{out}=10^{-20}$ g cm$^{-3}$, corresponding to negligible self-gravity ($\tilde{\beta}=1.3\times 10^{-7}$); the other corresponds to the critical solution with the maximum possible value of $\rho_\mathrm{out}$, beyond which the numerical calculation no longer converges. The maximum values of $\tilde{\beta}$ are 0.544, 0.654, and 0.868 for $\gamma=1$, $4/3$, and $1.66$, respectively. While these values are below $\beta_\mathrm{max} \approx 0.95$, they are derived using $\rho_\mathrm{out}$ rather than $\bar{\rho}$ (where $\bar{\rho}>\rho_\mathrm{out}$). The actual values of $\beta$, calculated with $\bar{\rho}$, are larger and could sometimes exceed $\beta_\mathrm{max}$.

The discrepancies between the solutions with negligible and critical values of $\tilde{\beta}$ are primarily evident in the outer regions, where most of the flow mass is distributed. For critical solutions, the profiles of $|v_r|$ become flat at the outer boundary, as theoretically anticipated (see Section \ref{upperlimit}).
In fact, ${\mathrm{d}\ln |v_r|}/{\mathrm{d}\ln r}$ can become slightly negative for some values of $\gamma$ (e.g., the bottom-left panel of Figure \ref{fig1}), though a further increase in $\beta$ would result in a rapid escalation of computational errors, ultimately leading to divergence in the numerical calculation. In solutions with negligible self-gravity, the density profiles are almost flat outside $r_\mathrm{B}$, as gas motion there is dominated by random thermal motion \citep{Frank2002}. By contrast, critical solutions exhibit dominance of self-gravity in the outer regions, which has to be balanced by the pressure gradient of the flow, since the flow is highly subsonic, making the radial acceleration term in Equation (\ref{eq_r_motion2}) negligible. Consequently, densities increase rapidly near the outer boundary as $r$ decreases to generate sufficient pressure gradients to balance the self-gravity. However, self-gravity decreases rapidly as $r$ decreases, as shown by the profiles of $M_\mathrm{t}$ in the right panels. The rapid increases in $\rho$ are thus no longer needed away from the outer boundary, and the density profiles become flat until reaching $r_\mathrm{B}$, within which the gravity of the central accretor begins to dominate. The pressure gradient also depends strongly on $\gamma$, which can be expressed as
\begin{equation}\label{eq_p_grad}
    \frac{1}{\rho} \frac{\mathrm{d} p}{\mathrm{d} r} = \gamma K{\rho}^{\gamma-1}\frac{\mathrm{d} \ln{\rho}}{\mathrm{d} r},
\end{equation}
where $K$ is a constant.
Therefore, larger $\gamma$ values reduce the required inward density increase to balance gravity, resulting in lower density profiles. As $\gamma$ approaches 5/3, the discrepancy between the density profiles of the two solutions becomes negligible in the inner regions, and the accretion rate remains almost unchanged even in the presence of strong self-gravity.

In deriving the analytical formulae in Section \ref{asol}, we adopted two key approximations: (1) the density of the flow is nearly constant outside $r_\mathrm{B}$; (2) $M_\mathrm{t}(r_\mathrm{c}) \approx M_*$. The second approximation holds for all solutions shown in Figure \ref{fig1} (see right panels), where the crosses represent the sonic points and $M_\mathrm{t}(r_\mathrm{c})/M_*$ is close to 1. The first approximation, however, only holds for small values of $\tilde{\beta}$. As $\tilde{\beta}$ approaches its upper limit, the steep increase in density near the outer boundary indicates that assuming a constant density in the outer regions could introduce noticeable inaccuracies. Quantitative analyses of this effect follow in subsequent sections.
\subsubsection{Solutions for Identical $\tilde{\beta}$ under Diverse Parameter Combinations}\label{same_beta}

In our analytical framework, self-gravity effects are characterized by $\beta$. To systematically verify key model predictions---specifically, that the enhancement of accretion rate [Equations~\eqref{eq_mdot2} and \eqref{eq_mdot_gamma1}] and shift in sonic radius [Equation~\eqref{eq_xc3}] due to self-gravity depend solely on $\beta$ and $\gamma$ for $x_\mathrm{out} \gg 1$---we examine five purpose-built parameter sets (Table \ref{tab1}). These are illustrative examples rather than representations of specific astrophysical systems. They are artificially chosen to span a representative range of conditions while maintaining $\tilde{\beta} = 0.65$ and $\gamma = 4/3$.

The parameter sets are constructed as follows. Case A serves as the baseline configuration: $M_* = M_\odot$, $a_\mathrm{out} = 10^{-4}c$, and $x_\mathrm{out} = 1000$ (corresponding to $r_\mathrm{out} = 2.95 \times 10^{16}$ cm), with $\rho_\mathrm{out} = 5.02 \times 10^{-14}$ g cm$^{-3}$ set via Equation~\eqref{eq_beta2} to ensure $\tilde{\beta}=0.65$. We denote these values as $M_\mathrm{A}$, $a_\mathrm{A}$, $r_\mathrm{A}$, and $\rho_\mathrm{A}$ for normalization. Cases B--D apply parameter scalings that preserve $\tilde{\beta}$: 
Case B scales outer radius and sound speed ($r_\mathrm{out}=3r_\mathrm{A}$ and $a_\mathrm{out}=3a_\mathrm{A}$); 
Case C scales gas density and sound speed ($\rho_\mathrm{out}=10\rho_\mathrm{A}$ and $a_\mathrm{out}=\sqrt{10}a_\mathrm{A}$); 
Case D scales gas density and outer radius ($\rho_\mathrm{out}=100\rho_\mathrm{A}$ and $r_\mathrm{out}=0.1r_\mathrm{A}$). 
Case E sets $M_* = 10M_\mathrm{A}$ while holding other parameters fixed, which also preserves $\tilde{\beta}$ since its final expression in Equation~\eqref{eq_beta2} is independent of $M_*$.

\begin{figure*}[ht]
    \centering
    \includegraphics[width=\textwidth]{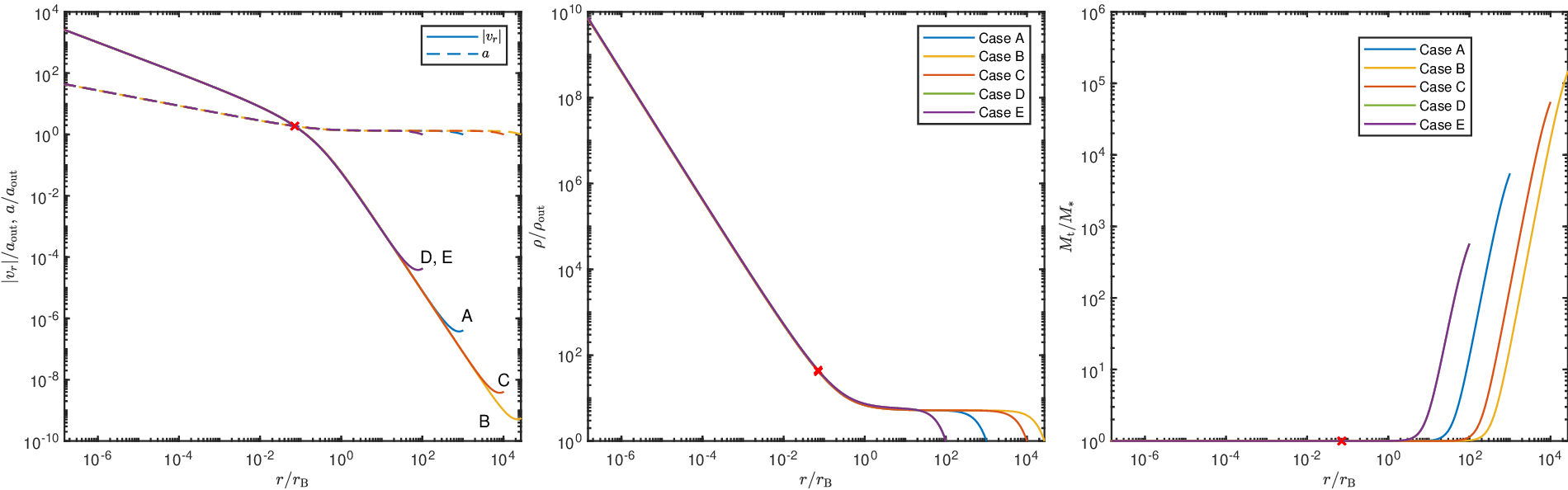}
    \caption{Global solutions of the TPBVP for $\tilde{\beta}=0.65$ and $\gamma=4/3$. The parameter settings in different Cases are listed in Table \ref{tab1}. The curves for Cases D and E coincide in each panel.}
    \label{fig2}
\end{figure*}

The resulting transonic global solutions for these five Cases are shown in Figure~\ref{fig2}, illustrating the impact of varied conditions under constant $\tilde{\beta}$. Crucially, normalized physical quantity profiles converge away from the outer boundary, indicating that while the absolute accretion rate may vary with parameters, the enhancement of the accretion rate due to self-gravity, $\dot{m}$, remains nearly constant.
The values of $x_\mathrm{c}$ and $\dot{m}$ are shown in Table \ref{tab1}, which exhibit close agreement for identical $\tilde{\beta}$. This confirms that $\tilde{\beta}$ effectively characterizes self-gravity effects in steady, spherically symmetric accretion.
\begin{table*}[ht]
  \centering
  \caption{Summary of solutions for $\tilde{\beta}=0.65$ and $\gamma=4/3$.}
  \label{tab1}
  \begin{tabular}{cccccccccccc}
    \hline\hline    
    {Case} & {$\rho_\mathrm{out}/\rho_\mathrm{A}$} & {$r_\mathrm{out}/r_\mathrm{A}$} & {$a_\mathrm{out}/a_\mathrm{A}$} & {$M_*/M_\mathrm{A}$} & {$\tilde{\beta}$} & {$x_\mathrm{out}$} & {$x_\mathrm{c}$} & {$\dot{m}$} & {$\beta$} & {$x_\mathrm{c,ana}$} & {$\dot{m}_\mathrm{ana}$} \\
    \hline
    A & 1    & 1   & 1           & 1  & 0.65 & 1000  & 0.0721 & 2.28 & 1.31 & 0.0859 & 1.76 \\
    B & 1    & 3   & 3           & 1  & 0.65 & 27000 & 0.0723 & 2.27 & 1.30 & 0.0860 & 1.75 \\
    C & 10   & 1   & $\sqrt{10}$ & 1  & 0.65 & 10000 & 0.0723 & 2.27 & 1.30 & 0.0860 & 1.75 \\
    D & 100  & 0.1 & 1           & 1  & 0.65 & 100   & 0.0702 & 2.38 & 1.36 & 0.0849 & 1.79 \\
    E & 1    & 1   & 1           & 10 & 0.65 & 100   & 0.0702 & 2.38 & 1.36 & 0.0849 & 1.79 \\
    \hline    
  \end{tabular}
  \tablecomments{Here $\rho_\mathrm{A}=5.02\times 10^{-14}$ g cm$^{-3}$, $r_\mathrm{A}=2.95\times 10^{16}$ cm (corresponding to $x_\mathrm{out}=1000$), $a_\mathrm{A}=10^{-4}c$, and $M_\mathrm{A}=M_\odot$ are the parameter values adopted in Case A. These illustrative values are chosen to yield $\tilde{\beta}=0.65$ via Equation~\eqref{eq_beta2}. The subscript "ana" denotes the values calculated using analytical formulae for the corresponding $\beta$.}
\end{table*}

We additionally present $\beta$ values in Table \ref{tab1}, calculated using the mean density $\bar{\rho}$ obtained from the global solutions. Although these values are not exactly identical, they exhibit close agreement for solutions corresponding to identical $\tilde{\beta}$, and $\dot{m}$ varies monotonically with $\beta$ even for slight $\beta$ changes. This shows that $\beta$ characterizes self-gravity effects more effectively than $\tilde{\beta}$, which is to be expected since $\tilde{\beta}$ is only an estimate of $\beta$.
As discussed in Section \ref{dif_beta}, the density profile beyond $r_\mathrm{B}$ deviates from near-constancy in the presence of strong self-gravity, leading to $\beta > \tilde{\beta}$. 
We note that the numerically obtained $\beta$ can exceed $\beta_\mathrm{max}$ predicted by the analytical formula, since numerical solutions allow slightly negative values of ${\mathrm{d}\ln |v_r|}/{\mathrm{d}\ln r}$ at the outer boundary. Table \ref{tab1} also lists the values of $x_\mathrm{c}$ and $\dot{m}$ calculated via Equations (\ref{eq_xc3}) and (\ref{eq_mdot2}) for the corresponding $\beta$, denoted as $x_\mathrm{c,ana}$ and $\dot{m}_\mathrm{ana}$, respectively, which roughly agree with the numerical results.
\subsection{Accretion Properties versus $\tilde{\beta}$ and Comparison with Analytical Predictions}\label{mdot_vs_beta}

\begin{figure}
\includegraphics[width=\columnwidth]{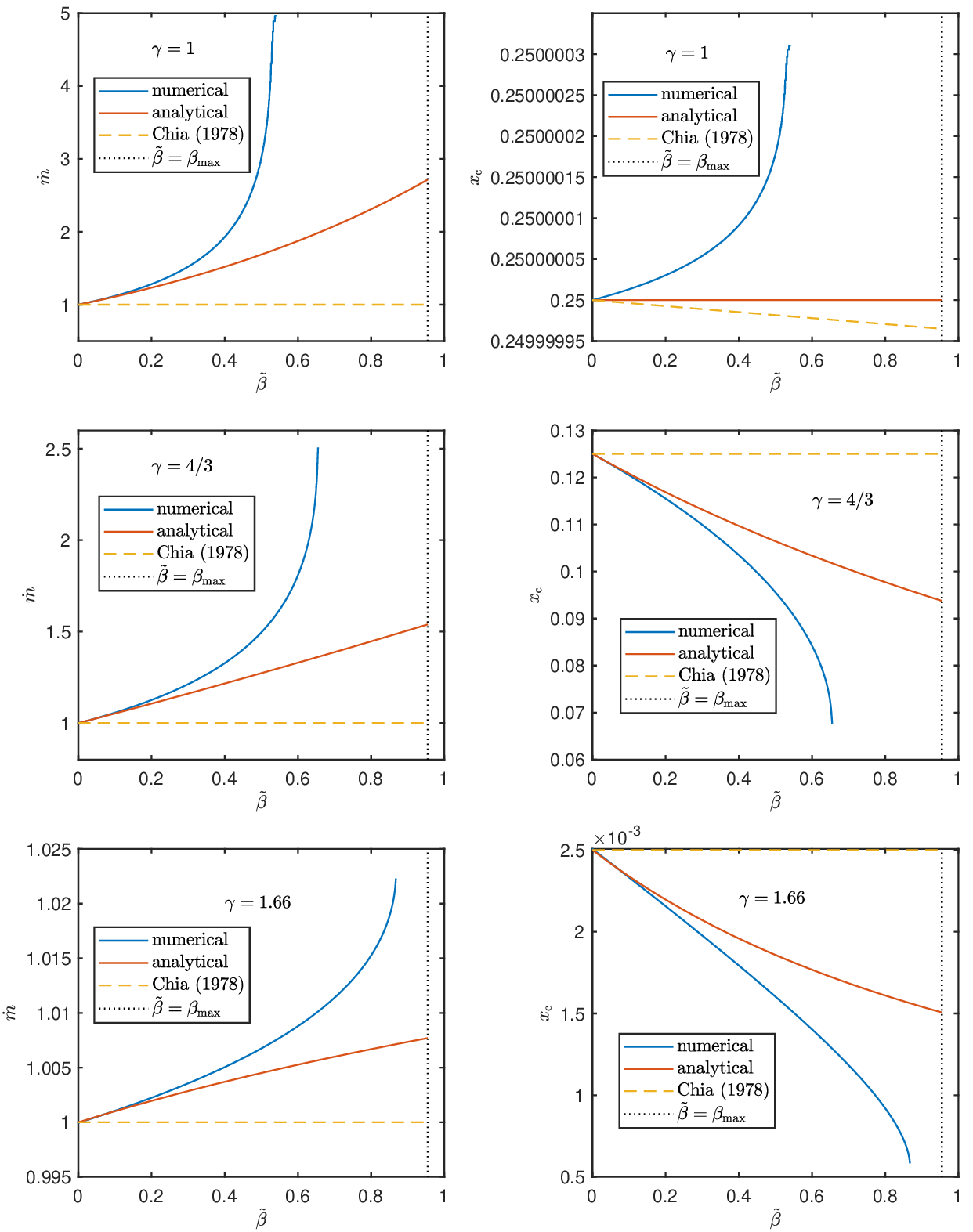}
\caption{Normalized accretion rate $\dot{m}$ (left column) and sonic radius $x_\mathrm{c}$ (right column) as functions of $\tilde{\beta}$. The top, middle, and bottom rows correspond to $\gamma = 1$, $4/3$, and $1.66$, respectively. Blue and red solid lines represent our numerical and analytical results, respectively. Black dotted lines mark the upper limits of $\tilde{\beta}$. Yellow dashed lines indicate the theoretical predictions of \citet{Chia1978}. }
\label{fig3}
\end{figure}

This section validates our analytical formulae for accretion properties---specifically the normalized accretion rate $\dot{m}$ and sonic radius $x_\mathrm{c}$---against numerical solutions of the TPBVP, and contrasts them with those of \citet{Chia1978}.
The $\tilde{\beta}$-dependence of $\dot{m}$ (left column) and $x_\mathrm{c}$ (right column) is presented in Figure \ref{fig3}, with
blue solid lines representing numerical solutions, red solid lines our analytical formulae, and yellow dashed lines the analytical predictions of \citet{Chia1978}.
Here we fix $a_\mathrm{out}=10^{-4}c$, $x_\mathrm{out}=1000$, and $M_*=M_\odot$, adjusting $\rho_\mathrm{out}$ to vary $\tilde{\beta}$. While $\beta$ is not equal to $\tilde{\beta}$, it increases monotonically with $\tilde{\beta}$. We aim to evaluate the validity of the analytical formulae when $\tilde{\beta}$ is used as an estimate of $\beta$.

Our analytical formulae agree with the numerical results qualitatively. Specifically, Equation (\ref{eq_mdot2}) predicts that $\dot{m}$ increases with increasing $\beta$ and decreases with increasing $\gamma$, approaching unity as $\gamma \to 5/3$. This is clearly shown in  Figure \ref{fig3}, with larger $\dot{m}$ for larger $\tilde{\beta}$ and smaller $\gamma$. For $\gamma=1.66$, $\dot{m}$ only reaches 1.02 at its numerical maximum, which is a negligible enhancement of the accretion rate due to self-gravity. Equation (\ref{eq_xc3}) predicts that $x_\mathrm{c}$ decreases with increasing $\beta$ for $1<\gamma<5/3$ and remains constant at 0.25 for $\gamma=1$. The decrease of $x_\mathrm{c}$ is clearly shown for $\gamma=4/3$ and 1.66. For $\gamma=1$, we have mentioned in Section \ref{ana_eqs} that $x_\mathrm{c}$ increases slightly as $\beta$ increases according to Equation (\ref{N1}), considering that $M_t(r_\mathrm{c})$ is actually slightly larger than $M_*$. This is also verified in the top-right panel of Figure \ref{fig3}.

Quantitatively, our analytical formulae show good agreement with numerical solutions at low self-gravity strengths. As $\tilde{\beta}$ increases, numerical results systematically exceed our analytical predictions for $\dot{m}$, with deviations growing to factors of as large as 2--3 (for $\gamma=1$) near the numerical upper limits of $\tilde{\beta}$. This divergence stems from two limitations of our analytical approach: (1) the approximation $\bar{\rho} \approx \rho_\mathrm{out}$ becomes increasingly invalid with increasing $\tilde{\beta}$; (2) the assumed constant density profile cannot capture the steep radial gradients induced by self-gravitational compression at the outer boundary (see Figures \ref{fig1} and \ref{fig2}, middle panels).
The analytical values of $\dot{m}_\mathrm{max}$ from Equation (\ref{eq_mdotmax}), indicated by the intersection points of the red solid lines and the black dotted lines in the left panels of Figure \ref{fig3}, are also lower than the numerical results. However, $\dot{m}_\mathrm{max}$ obtained from numerical solutions does not exceed twice that predicted by the analytical formula, even in the extreme case of $\gamma=1$. Therefore, Equation (\ref{eq_mdotmax}) can still be used as a rough estimate of the maximum enhancement of the accretion rate due to self-gravity.

According to Equations~\eqref{alpha} and \eqref{eq_beta2}, the self-gravitational parameter $\alpha$ introduced by \citet{Chia1978} relates to $\tilde{\beta}$ as
\begin{equation}\label{alpha_chia}
\alpha=\frac{\pi \tilde{\beta}}{2x_\mathrm{out}^2}.
\end{equation}
Under our parameter settings, $\alpha \ll 1$ and $r_* \ll r_\mathrm{B}$. In this regime, the self-gravity-corrected normalized accretion rate and sonic radius derived by \cite{Chia1978} can be expressed in our notation as
\begin{equation}\label{eq_mdot_chia}
    \dot{m} = \frac{\dot{M}}{\dot{M}_\mathrm{B}}=1+ \frac{\alpha}{12} \cdot\left(\frac{5-3\gamma}{2}\right)^{\frac{3\gamma-4}{\gamma -1}},
\end{equation}
\begin{equation}\label{eq_xc_chia}
x_\mathrm{c} = \frac{r_\mathrm{c}}{r_\mathrm{B}} = \left[1- \frac{\alpha(9\gamma-7)}{48(5-3\gamma)}\cdot \left(\frac{5-3\gamma}{2}\right)^{\frac{3\gamma-4}{\gamma -1}}\right]\cdot \frac{5-3\gamma}{8}.
\end{equation}
These expressions indicate that the self-gravity corrections derived by \citet{Chia1978} are linear functions of $\alpha$. 
Consequently, their effects are negligible in Figure~\ref{fig3}, where $x_\mathrm{out} = 1000$ leads to negligibly small $\alpha$ values according to Equation~\eqref{alpha_chia}.
The sole exception appears in the sonic radius profile for $\gamma = 1$ (top-right panel), where numerical solutions, our analytical formula, and that of \citet{Chia1978} all yield variations in $x_\mathrm{c}$ below $10^{-6}$. Despite being negligibly small in magnitude, the corrections predicted by \citet{Chia1978} become visually discernible in the plot due to the chosen vertical scale.

The fundamental difference between our analytical framework and that of \citet{Chia1978} lies in the treatment of the self-gravity potential.
Their formulation neglects all $r_\mathrm{out}$-dependent terms in the self-gravity potential (see Section~\ref{ana_eqs}), whereas ours includes the full potential.
Because our numerical framework directly solves the fluid equations without assuming a specific potential form, it provides an independent test of the analytical results.  Figure~\ref{fig3} shows that the numerical results are more consistent with our analytical formulae than with those of \citet{Chia1978}.

The dependence of self-gravity effects on the outer radius ($r_{\mathrm{out}}$, hence $x_\mathrm{out}$) of the accretion flow---a key driver of the divergence between the two models in Figure~\ref{fig3}---constitutes a critical distinction that persists across a broader parameter space.
Here, $r_{\mathrm{out}}$ represents the characteristic length scale of the ambient gas cloud when unconstrained by external influences (e.g., tidal forces; see Section~\ref{app2}).
For fixed gas density and sound speed, \citet{Chia1978} predicts self-gravity effects independent of $r_{\mathrm{out}}$---a feature that holds generally in their formulation, not limited to the small-$\alpha$ approximation.
In contrast, our results show that self-gravity strengthens significantly with increasing $r_{\mathrm{out}}$, since $\beta \propto r_\mathrm{out}^2$ according to Equation~\eqref{eq_beta}, and ultimately destabilizes the accretion flow when $\beta$ exceeds a critical threshold [Equation~\eqref{eq_ulbeta}].
This trend is consistent with the fundamental theory of gravitational collapse. Classical criteria---such as the Jeans length or the more refined Bonnor-Ebert threshold, which accounts for density stratification and external pressure confinement---predict that collapse occurs once the system size exceeds a critical value \citep{Ebert1955,Bonnor1956,book2017}. Crucially, both observations and numerical simulations \citep[][and references therein]{Enr2019,Tra2020,Moon2024} have confirmed that self-gravitating gas clouds indeed collapse beyond specific size thresholds. 
The agreement between our results and these theoretical and empirical studies validates our self-gravity formulation, which incorporates the full gravitational potential of the gas. 
Conversely, the model of \citet{Chia1978} fails to predict such collapse because it neglects all $r_\mathrm{out}$-dependent terms in the self-gravity potential.
\section{Applications to Astrophysical Scenarios}\label{apps}
In this section, we examine two astrophysical scenarios in which self-gravity may play a significant role: (1) hyper-Eddington accretion onto SMBH seeds in the early Universe, and (2) accretion onto SMOs embedded in AGN disks. For each case, we estimate realistic outer boundary conditions based on the expected physical environments and assess the significance of self-gravity accordingly.

\subsection{Hyper-Eddington Accretion onto Supermassive Black Hole Seeds in the Early Universe}\label{app1}

Observations of SMBHs in the early Universe present a significant challenge to classical accretion theory. One prominent example is the quasar J0313$-$1806 \citep{WFG2021}, which hosts a SMBH of mass $M_\mathrm{SMBH}=1.6 \times 10^9 M_\odot$ at redshift $z=7.642$.
Assuming that a seed BH of mass $M_\mathrm{seed}$ accretes at the Eddington limit, $\dot{M}_\mathrm{Edd}=L_\mathrm{Edd}/(\eta c^2)$, the growth time to $M_\mathrm{SMBH}$ is given by \citep{Kohei2020}
\begin{equation}\label{t_grow}
t_{\mathrm{grow}} \approx \frac{0.45 \eta}{(1-\eta) f_{\mathrm{duty }}} \ln \left(\frac{M_\mathrm{SMBH}}{M_{\mathrm{seed }}}\right)\, \mathrm{Gyr},
\end{equation}
where $\eta$ is the radiative efficiency and $f_{\mathrm{duty }}$ is the duty cycle (the fraction of time the BH is actively accreting). For fiducial parameters of $\eta=0.1$, $f_{\mathrm{duty }}=1$, and $M_\mathrm{seed}=100 M_\odot$, this yields $t_{\mathrm{grow}} \approx 0.83$ Gyr, exceeding the age of the Universe at $z=7.642$ ($\approx 0.68$ Gyr in the $\Lambda$CDM model; \citealt{Planck2020}).
A natural resolution to this tension invokes substantially more massive seed BHs in the early Universe, potentially formed either through rapid gas accretion onto a stellar-mass BH, or via the direct collapse of chemically pristine primordial gas in atomic-cooling halos \citep[see][for a review of these and alternative explanations]{Kohei2020}. 
The former scenario is supported by numerical simulations of \cite{Kohei2016}, which demonstrate that under sufficiently high ambient gas densities, the inner region of a spherically symmetric accretion flow forms a core dominated by photon trapping, while the outer region (beyond $10^{-3} r_\mathrm{B}$) maintains the classical Bondi profile.  
The final steady state of the accretion flow in their simulations resembles an isothermal Bondi flow, with an accretion rate $\dot{M} \simeq \dot{M}_{\mathrm{B}} \gtrsim 5000 L_{\mathrm{Edd}}/c^2$, orders of magnitude higher than $\dot{M}_\mathrm{Edd}$ for $\eta=0.1$. 
This hyper-Eddington accretion can sustain until the BH mass reaches $\sim 1.4 \times 10^5 M_\odot$, rapidly forming a massive seed BH.
In their calculation, a stellar-mass BH is embedded in a dense gas cloud at the center of a dark matter halo, whose density profile is given by
\begin{equation}\label{n_halo}
n(r) \simeq 10^3 T_{\mathrm{vir}, 4} \mathrm{~cm}^{-3}\left(\frac{r}{10 \mathrm{pc}}\right)^{-2}\left(\frac{f_{\mathrm{n}}}{4}\right),
\end{equation}
where $f_\mathrm{n}=4$ and $T_{\mathrm{vir}, 4}$ is the virial temperature of the halo normalized by $10^4$ K. The $r^{-2}$ dependence follows cosmological simulations of high-$z$ protogalaxies \citep[e.g.,][]{Wise2008,Regan2014}. The Bondi accretion rate in this scenario is found to be above 0.1 $M_\odot$ yr$^{-1}$, varying with the mass and virial temperature of the halo. 
A stellar-mass BH can thus grow to $\sim 10^5 M_\odot$ within 1 Myr. 
Using this seed mass in Equation (\ref{t_grow}) yields $t_{\mathrm{grow}} \approx 0.48$ Gyr to reach $1.6 \times 10^9 M_\odot$, slightly shorter than the cosmic time between $z=20$ and $z=7.642$ ($\approx 0.50$ Gyr).

We estimate the impact of self-gravity during the hyper-Eddington accretion phase using the dimensionless parameter $\tilde{\beta}$. Substituting the gas density profile from Equation~\eqref{n_halo} into Equation~\eqref{eq_beta2} yields
\begin{equation}\label{beta_app1}
\tilde{\beta} = \frac{2G\rho_\mathrm{out} r_\mathrm{out}^2}{a_\mathrm{out}^2}
= \frac{2G \mu m_\mathrm{p} n(r_\mathrm{out}) r_\mathrm{out}^2}{\gamma R_\mathrm{gas} T_\mathrm{vir}/\mu} \approx 0.38,
\end{equation}
where $\gamma = 1$, $\mu = 1.22$ \citep{Kohei2016}, and $m_\mathrm{p}$ is the proton mass.
This corresponds to a $50\%$ enhancement of the accretion rate due to self-gravity, as predicted by  Equation~\eqref{eq_mdot_gamma1}.
Such an enhancement is non-negligible, as it permits greater flexibility in model parameters (e.g., the duty cycle during this phase) while maintaining consistency with the timescale required for seed BH growth.
Moreover, the relatively large value of $\tilde{\beta} = 0.38$ indicates that self-gravity plays a significant role in hyper-Eddington accretion onto SMBH seeds, and should therefore be incorporated into future theoretical models and numerical simulations. This inclusion may yield quantitatively distinct outcomes that more accurately capture the underlying physics.

\subsection{Accretion onto Stellar-Mass Objects Embedded in AGN Disks}\label{app2}

AGN disks provide extreme environments where embedded SMOs, including stars and compact objects, undergo hyper-Eddington accretion \citep{McK2012,Cant2021,Li2021}.  
Studies of the accretion process typically assume that the SMO corotates with the AGN disk and adopt a non-inertial frame corotating with it. In this frame, the relative motion between the SMO and the surrounding gas can be neglected (e.g., \citealt{Wang2021,Wang2021b}; but see \citealt{Kocsis2011} for discussions including relative motion). The forces acting on the gas include the gravitational force of the SMO and the tidal force from the central SMBH (the combination of the centrifugal force and the gravitational force of the SMBH). The Coriolis force is also present but is commonly omitted in simplified treatments of this scenario, as in previous studies. We follow this approximation here.

A standard characteristic scale in the corotating frame is the Hill radius \citep{book2015}, 
\begin{equation}\label{rh1}
r_\mathrm{H} \equiv \left( \frac{M_2}{3M_1} \right)^{1/3} R,
\end{equation}
where $M_1$ and $M_2$ denote the masses of the SMBH and the SMO, respectively, and $R$ is their separation. 
The Hill radius demarcates the boundary at which the tidal force from the SMBH balances the gravitational attraction of the SMO.
When the self-gravity of the gas is taken into account, the gravitational attraction from the gas enclosed within $r_\mathrm{H}$ also contributes to offsetting the tidal force. 
In this case, the effective Hill radius (denoted $r_\mathrm{H,sg}$) is modified and satisfies the following balance:
\begin{equation}\label{eq_rh2}
3\Omega^2 r_\mathrm{H,sg} = \frac{G}{r_\mathrm{H,sg}^2} \left( M_2 + \frac{4\pi}{3} r_\mathrm{H,sg}^3 \rho_\mathrm{d} \right),
\end{equation}
where $\Omega=\sqrt{GM_1/R^3}$ is the orbital angular velocity of the SMO around the SMBH, $\rho_\mathrm{d}$ is the local gas density, and the term $4\pi r_\mathrm{H,sg}^3\rho_\mathrm{d}/3$ corresponds to the mass of gas enclosed within $r_\mathrm{H,sg}$. The left-hand side represents the tidal acceleration from the SMBH.
Solving for $r_\mathrm{H,sg}$, we obtain
\begin{equation}\label{rh2}
r_\mathrm{H,sg} \equiv \left( \frac{M_2}{3M_1-4\pi\rho_\mathrm{d}R^3/3} \right)^{1/3} R,
\end{equation}
which reduces to Equation~\eqref{rh1} in the limit $\rho_\mathrm{d} \to 0$.
Within $r_\mathrm{H,sg}$, the combined gravity of the SMO and its bound gas dominates, while outside this radius, the tidal force from the SMBH dominates. As a rough approximation, we take $r_\mathrm{H,sg}$ as the outer boundary of the accretion region of the SMO and neglect the tidal influence of the SMBH within this radius.
We note that in certain regimes, $r_\mathrm{H,sg} < r_\mathrm{B}$, where $r_\mathrm{B}$ is the Bondi radius of the SMO. 
Under such conditions, the analytical formula may yield $\dot{M} < \dot{M}_\mathrm{B}$---even in the presence of non-negligible self-gravity---due to a reduced accretion region.
Moreover, the discrepancy between analytical predictions and numerical solutions becomes more significant in these regimes, since the assumption $x_\mathrm{out} \gg 1$ is no longer valid. 
In what follows, we examine the cases $r_\mathrm{H,sg} > r_\mathrm{B}$ and $r_\mathrm{H,sg} < r_\mathrm{B}$ separately. For the latter case, we additionally calculate numerical solutions to characterize the accretion flow.

We consider an AGN disk surrounding a SMBH with mass $M_1=10^8M_\odot$, accreting at a rate $\dot{M}_1=L_\mathrm{Edd1}/c^2$, where $L_\mathrm{Edd1} = {4\pi c G M_1 }/{\kappa_\mathrm{es}}$ is the Eddington luminosity of the SMBH and $\kappa_\mathrm{es}$ denotes the electron scattering opacity. 
The disk structure is obtained by numerically solving the equations of the standard thin disk model \citep[SSD;][]{SS73,Kato2008}, adopting a viscosity parameter $\alpha=0.1$ and assuming solar abundances (with mean molecular weight $\mu=0.6$ and $\kappa_\mathrm{es}=0.35$ cm$^2$ g$^{-1}$). 
As mentioned in Section~\ref{app1}, numerical simulations by \cite{Kohei2016} show that the steady state of a hyper-Eddington, spherically symmetric accretion flow resembles an isothermal Bondi flow. 
We therefore adopt an adiabatic index $\gamma = 1$ in our calculations. 
The resulting accretion rate is then checked against the hyper-Eddington criterion $\dot{M} \gtrsim 5000 L_\mathrm{Edd}/c^2$ from \citet{Kohei2016} to verify the consistency of this assumption.
The SMO mass is set to $M_2=M_\odot$.  

At $R=10R_\mathrm{S}$, where $R_\mathrm{S}=2GM_1/c^2$ is the Schwarzschild radius of the SMBH, we obtain values of $r_\mathrm{B}=8.0 \times 10^{10}$ cm, $r_\mathrm{H,sg}=4.4 \times 10^{11}$ cm, 
$\rho_\mathrm{d}=1.5\times 10^{-8}$ g cm$^{-3}$, and $a_\mathrm{d}=5.8\times 10^7$ cm s$^{-1}$. 
Here, $a_\mathrm{d}$ denotes the local sound speed of the disk gas, and together with $\rho_\mathrm{d}$, they are used as the outer boundary conditions $\rho_\mathrm{out}$ and $a_\mathrm{out}$ in our calculations.
Substituting these values into Equation \eqref{eq_beta2} yields $\tilde{\beta}=1.2\times 10^{-7}$, indicating that self-gravity is negligible in this case. 
From Equation \eqref{eq_mdot_gamma1}, with a negligible $\tilde{\beta}$ and $x_\mathrm{out}=r_\mathrm{H,sg}/r_\mathrm{B}=5.5$, we get the accretion rate in this case,
$\dot{M}=0.91\dot{M}_\mathrm{B}=1.1 \times 10^5 L_\mathrm{Edd2}/c^2$, where $L_\mathrm{Edd2} = {4\pi c G M_2 }/{\kappa_\mathrm{es}}$ is the Eddington luminosity of the SMO.\footnote{Note that $\dot{M}<\dot{M}_\mathrm{B}$ is expected, as $\dot{M}_\mathrm{B}$ corresponds to the classical Bondi solution where the outer boundary is located at infinity, while our model adopts a finite $r_\mathrm{out}$ (see Section~\ref{asol}).} This confirms that the accretion flow is hyper-Eddington.

By contrast, at $R=1000R_\mathrm{S}$, we obtain values of $r_\mathrm{B}=9.3 \times 10^{13}$ cm, $r_\mathrm{H,sg}=4.8 \times 10^{13}$ cm, $\rho_\mathrm{d}=1.3\times 10^{-9}$ g cm$^{-3}$, and $a_\mathrm{d}=1.7\times 10^6$ cm s$^{-1}$. Substituting these values into Equation \eqref{eq_beta2} yields $\tilde{\beta}=0.14$, indicating that self-gravity is non-negligible.
From Equation \eqref{eq_mdot_gamma1}, the accretion rate without self-gravity is $\dot{M}_0=0.38\dot{M}_\mathrm{B}$, corresponding to $\beta=0$ and $x_\mathrm{out}=r_\mathrm{H,sg}/r_\mathrm{B}=0.52$; the accretion rate with self-gravity is $\dot{M}_\mathrm{sg}=0.44\dot{M}_\mathrm{B}$, corresponding to $\beta=0.14$ and the same value of $x_\mathrm{out}$.
More accurate numerical solutions yield $\dot{M}_0=0.42\dot{M}_\mathrm{B}$ and $\dot{M}_\mathrm{sg}=0.54\dot{M}_\mathrm{B}$, showing a $28\%$ accretion rate enhancement due to self-gravity. The corresponding Bondi accretion rate is $\dot{M}_\mathrm{B}=4.1 \times 10^8 L_\mathrm{Edd2}/c^2$, confirming that the flow is hyper-Eddington.

A comparison between the two cases suggests that self-gravity effects become more significant at larger separation $R$.
Although the exact dependence of $\tilde{\beta}$ on $R$ is complex due to simultaneous changes in disk gas properties, the increase in the outer radius $r_\mathrm{out} = r_\mathrm{H,sg}$ [Equation~\eqref{rh2}] generally dominates, resulting in  larger $\tilde{\beta}$ at greater $R$.
Similarly, self-gravity strengthens with increasing $M_2$, which also enlarges $r_\mathrm{H,sg}$.
These trends indicate that non-negligible self-gravity is likely to arise in a broad range of such accretion scenarios.

In summary, self-gravity can play a significant role in SMO accretion within AGN disks.
Its importance depends on local disk conditions, the masses of the SMO and the SMBH, and their separation, and should be assessed using the dimensionless parameter $\tilde{\beta}$.
When self-gravity becomes significant, it enhances the total gravitational field and increases the effective Hill radius, thereby expanding the size of the accretion region and altering its structure.

\section{Summary and discussion}\label{summary}
We investigate steady, spherically symmetric accretion flows with self-gravity using both analytical and numerical methods. In the analytical framework, self-gravity effects can be approximately characterized by a dimensionless parameter $\beta$. The influence of self-gravity on the structure of the accretion flow also depends on the adiabatic index $\gamma$ and, to a lesser degree, on the normalized outer radius $x_\mathrm{out}$. 
The accretion flow approaches the classical Bondi solution as $\beta \to 0$. For $1<\gamma<5/3$, the sonic radius shifts inward and the accretion rate is enhanced as $\beta$ increases. This enhancement, quantified by $\dot{m}$, becomes more pronounced as $\gamma$ decreases. For $\gamma=1$, the accretion rate experiences the strongest enhancement, while the sonic radius remains nearly constant at $x_\mathrm{c}=0.25$. Conversely, for $\gamma=5/3$, the sonic radius is always at the origin and the accretion rate shows no enhancement due to self-gravity. 
We also find an upper limit for $\beta$, beyond which steady, spherically symmetric accretion becomes unsustainable. The upper limit exhibits remarkable agreement with constraints from gravitational instability criteria, despite originating from fundamentally different physical formulations.
In the numerical approach, transonic global solutions are obtained by directly solving the TPBVP with BCs at the outer boundary, the sonic point, and the accretor surface. Global solutions for identical $\tilde{\beta}$ (an estimate of $\beta$) exhibit rapid convergence interior to $r_\mathrm{out}$, demonstrating the efficacy of $\tilde{\beta}$ in characterizing self-gravity effects. Systematic parameter surveys confirm the analytical predictions described above. Numerical solutions also encounter an upper limit for $\tilde{\beta}$, beyond which the numerical algorithm fails to converge. The values of $\beta_\mathrm{max}$, calculated with $\bar{\rho}$ obtained from the global solutions, could exceed those predicted by the analytical formula (see Table \ref{tab1}). 
The approximate analytical formulae for $\dot{m}$ and $x_\mathrm{c}$ agree well with the numerical results for small $\tilde{\beta}$, though the discrepancy grows with increasing $\tilde{\beta}$. The maximum accretion rate predicted by the analytical formula lies in the range of 50--100\% of that obtained from the numerical solution, providing a rough estimate of $\dot{m}_\mathrm{max}$ without solving the TPBVP.

We also compare our analytical formulae with those of \citet{Chia1978}.
The fundamental difference between the two frameworks lies in the treatment of the self-gravity potential.
\citet{Chia1978} neglects all terms in the self-gravity potential that depend on the outer radius $r_\mathrm{out}$, which characterizes the size of the ambient gas cloud when unconstrained by external influences (e.g., tidal forces). As a result, their predictions are independent of $r_\mathrm{out}$.
In contrast, our analysis incorporates the full self-gravity potential and predicts that the self-gravity strength, quantified by $\beta$, scales as $r_\mathrm{out}^2$.
We evaluate these differing predictions in two ways.
First, our numerical framework directly solves the fluid equations without assuming a specific potential form, thereby providing an independent test of the analytical results. As shown in Figure~\ref{fig3}, the numerical results agree more closely with our analytical formulae than with those of \citet{Chia1978}.
Second, both theoretical and empirical studies of gravitational instability show that self-gravitating gas clouds collapse when their size exceeds a critical threshold. This scale-dependent instability is consistent with our prediction that self-gravity strengthens with increasing $r_\mathrm{out}$ and ultimately destabilizes the accretion flow when $\beta$ exceeds a critical value.

When accounting for temporal evolution, Equation (\ref{eq_dlnvr}) generalizes to
\begin{equation}\label{eq_dlnvr2}
	\frac{r}{a^2}\frac{\partial v_r}{\partial t}	+ \left(\frac{v_r^2}{a^2}-1\right) \frac{\partial\ln |v_r|}{\partial\ln r}=2-\frac{G M_\mathrm{t}}{a^2r}.
\end{equation}
The right-hand side (RHS) becomes negative for self-gravity exceeding $\beta_\mathrm{max}$, and we expect that ${\mathrm{d}\ln |v_r|}/{\mathrm{d}\ln r}<0$ for an accretion flow. Initially, the flow should be subsonic ($v_r^2<a^2$) at the outer boundary, and we get ${\partial v_r}/{\partial t}<0$. Since $v_r<0$ for inflow, ${\partial v_r}/{\partial t}<0$ means that the magnitude of $v_r$ increases with time, and the flow is time-dependent. The structure of such flows requires time-dependent analyses, such as numerical simulations, which are beyond the scope of this paper.

In this paper, we adopt the adiabatic assumption of the classical Bondi model. However, radiative cooling may become non-negligible at high densities. Numerical simulations by \cite{Kohei2016} reveal that under such conditions, the inner region of a spherically symmetric flow forms a core dominated by photon trapping, while the outer region (beyond $10^{-3} r_\mathrm{B}$) maintains the classical Bondi profile. This spatial decoupling implies that the adiabatic approximation remains valid in regimes where self-gravity is dynamically significant. A detailed investigation incorporating radiative cooling will be conducted in future work.

Our analytical and numerical investigations of self-gravitating spherical accretion provide significant insights into astrophysical systems where self-gravity is non-negligible. 
To demonstrate its utility, we apply our formalism to two representative scenarios: (1) the hyper-Eddington accretion phase of SMBH seeds in the early Universe, and (2) accretion onto SMOs embedded in AGN disks. 
In both cases, we demonstrate that self-gravity can significantly enhance the accretion rate. 
For early-Universe SMBH seeds, this enhancement permits greater flexibility in model parameters while maintaining consistency with the observationally constrained growth time of seed BHs.
In AGN disks, self-gravity also modifies the effective Hill radius of embedded objects, consequently altering the size and structure of their accretion regions. 
These findings highlight the necessity of accounting for self-gravity when modeling dense astrophysical environments and demonstrate the utility of our analytical formulae in providing rough estimates without requiring full numerical solutions.
\begin{acknowledgments}
    We thank the anonymous reviewer for their constructive comments and valuable suggestions that helped improve the manuscript.
    This work was supported by the International Cooperation Projects of the National Key R\&D Program (No. 2022YFE0116800), the Science Foundation of Yunnan Province (No. 202401AS070046), the International Partnership Program of Chinese Academy of Sciences (No. 020GJHZ2023030GC) and the Yunnan Revitalization Talent Support Program.
\end{acknowledgments}

\bibliography{refs}{}
\bibliographystyle{aasjournal}

\end{CJK*}
\end{document}